\documentclass[final,5p,times,twocolumn,nopreprintline]{elsarticle}
\usepackage{amssymb}
\usepackage{amsmath}
\usepackage{xcolor}
\journal{Computer Physics Communications}

\begin{document}
	
\begin{frontmatter}
    
    \title{Path-integral molecular dynamics with actively-trained and universal machine learning force fields}
    
    \author[MSU,SINP,SKOLT]{A.A. Solovykh\corref{cor}}
    \author[SKOLT,DMLLC]{N.E. Rybin}
    \author[HSE,SKOLT,MIPT,EIBPRAS]{I.S. Novikov}
    \author[SKOLT,DMLLC]{A.V. Shapeev}
    
    \cortext[cor]{Al.Solovykh@skoltech.ru}
    
    \affiliation[MSU]{organization={Lomonosov Moscow State University, Faculty of Physics},
        addressline={Leninskie gory 1(2)}, 
        city={Moscow},
        postcode={119991}, 
        country={Russian Federation}}
    
    \affiliation[SINP]{organization={Skobeltsyn Institute of Nuclear Physics, Lomonosov Moscow State University},
        addressline={Leninskie gory 1(2)}, 
        city={Moscow},
        postcode={119991}, 
        country={Russian Federation}}
    
    \affiliation[SKOLT]{organization={Skolkovo Institute of Science and Technology},
        addressline={Bolshoy boulevard 30}, 
        city={Moscow},
        postcode={121205}, 
        country={Russian Federation}}
    
    \affiliation[DMLLC]{organization={Digital Materials LLC},
        addressline={Kutuzovskaya St. 4a}, 
        city={Odintsovo},
        postcode={143001}, 
        country={Russian Federation}}

    \affiliation[HSE]{organization={HSE University,         Faculty of Computer Science},
        addressline={Pokrovsky boulevard 11},
        state={Moscow},
        postcode={109028}, 
        country={Russian Federation}}

    \affiliation[MIPT]{organization={Moscow Institute of Physics and Technology, Institutsky lane 9, Dolgoprudny, Moscow region, 141700, Russian Federation}}

    \affiliation[EIBPRAS]{Emanuel Institute of Biochemical Physics of the Russian Academy of Sciences, 4 Kosygin Street, Moscow, 119334, Russian Federation}
    
    %% Abstract
    \begin{abstract}
        Accounting for nuclear quantum effects (NQEs) can significantly alter material properties at finite temperatures. Atomic modeling using the path-integral molecular dynamics (PIMD) method can fully account for such effects, but requires computationally efficient and accurate models of interatomic interactions. Empirical potentials are fast but may lack sufficient accuracy, whereas quantum-mechanical calculations are highly accurate but computationally expensive. Machine-learned interatomic potentials offer a solution to this challenge, providing near-quantum-mechanical accuracy while maintaining high computational efficiency compared to density functional theory (DFT) calculations. In this context, an interface was developed to integrate moment tensor potentials (MTPs) from the MLIP-2 software package into PIMD calculations using the i-PI software package. This interface was then applied to active learning of potentials and to investigate the influence of NQEs on material properties, namely the temperature dependence of lattice parameters and thermal expansion coefficients, as well as radial distribution functions, for lithium hydride (LiH) and silicon (Si) systems. The results were compared with experimental data, quasi-harmonic approximation calculations, and predictions from the universal machine learning force field MatterSim. These comparisons demonstrated the high accuracy and effectiveness of the MTP-PIMD approach.
    \end{abstract}
    
    
    %    \item An interface was developed to utilize Moment Tensor Potentials (MTPs) in path-integral molecular dynamics (PIMD) calculations
    %    \item The influence of nuclear quantum effects on the linear thermal expansion (LTE) coefficients of  LiH and Si is investigated
    %    \item A high degree of agreement between the results of the MTP-PIMD approach and the Mattersim-PIMD approach and experimental data was found
    %    \item Using the previously mentioned approaches, the presence of a negative LTE coefficient for Si was detected  
    
    %% Keywords
    %\begin{keyword}
        
        
        
    %\end{keyword}
    
\end{frontmatter}

    
    
    \section{Introduction}
    \label{Intro}

    Accounting for nuclear quantum effects (NQEs) in atomistic modeling is crucial for accurately describing many phenomena including lattice thermal expansion~\cite{kim2018nuclear}, proton tunneling in molecules~\cite{litman2020multidimensional}, lattice dynamics of molecular crystals~\cite{polymorphs, phasehighpressure}, as well as many other properties of materials~\cite{markland2018_NQE, rossi2021_review}. The standard method for incorporating NQEs is the path-integral (PI) formalism~\cite{feynman1966quantum, khandekar2002path, zinn2010path}, which maps quantum statistical mechanics onto classical statistical mechanics of a ring polymer composed of multiple replicas of the system connected by harmonic springs. The effective classical PI partition function of this ring polymer system can be sampled using molecular dynamics (MD), leading to the path-integral molecular dynamics (PIMD) method~\cite{PIMD, Tuckerman_PIMD, Althorpe_PIMD}. This method has been implemented in the i-PI software package~\cite{ipi3.0}, which is responsible for the propagation of the nuclear dynamics and requires separate software packages to calculate the interaction between atoms.

    A major challenge associated with the description of interatomic interactions lies in striking the right balance between the accuracy of the calculations and computational cost. While classical empirical models typically lack the required accuracy, quantum mechanical methods such as density functional theory (DFT) demand high computational resources. The situation has been improved due to the emergence of machine-learning interatomic potentials (MLIPs), which can achieve near-quantum-mechanical accuracy with a significant increase in computational efficiency compared to DFT calculations \cite{MLIPs-rev1, MLIPs-rev2, MLIP-rev3}. In this study, we employed one of the MLIPs, specifically Moment Tensor Potential (MTP) \cite{Shapeev2016-mtp}, and the algorithm for active learning (AL) of MTP \cite{podryabinkin2017-AL, gubaev2018machine} designed to automatically construct a training set. MTP and algorithm for its AL are implemented in the MLIP-2 software package \cite{novikov2020mlip}. MTPs are applicable in various computational materials science topics, including crystal structure prediction~\cite{gubaev2019-csp, podryabinkin2019-csp, rybin2025accelerating}, lattice thermal conductivity calculations \cite{mortazavi2021-ltc, rybin2024moment}, hardness evaluation \cite{podryabinkin2022-nanohardness}, calculations of the self-diffusion coefficient of atoms \cite{novoselov2019-diffusion}, and the physico-chemical properties of melts~\cite{rybin2024thermophysical}. Recently, MTPs with explicit incorporation of magnetic degrees of freedom were developed~\cite{Novikov2022-mMTP} and applied to a variety of problems~\cite{Kotykhov2023-cDFT-mMTP, Kotykhov2024_mag_force_fit, Kotykhov2024-CrN}.   

    Previously, the method for AL of MTP (AL-MTP) was combined with another method describing NQEs, namely, ring polymer molecular dynamics (RPMD) \cite{craig2004-rpmd, craig2005-rpmd-refined, craig2005-rpmd, braams2006-rpmd}. This method was implemented in the RPMDrate code \cite{suleimanov2013-rpmdrate} and enables accurate calculations of chemical reaction rates. In \cite{novikov2018-rpmd-al-mtp,novikov2019-s+h2,novikov2024-oh+hbr}, it was demonstrated that the combination of AL-MTP and RPMD yields accurate rates for different chemical reactions. Moreover, MLIPs and training sets were created automatically during RPMD simulations. 
    
    Other MLIPs, such as the Behler–Parrinello neural network~\cite{BPNN}, MACE~\cite{MACE}, n2p2~\cite{n2p2_i-pi}, and DeePMD~\cite{DeePMD, wang2018deepmd}, have been successfully used in PIMD simulations. However, despite MTPs' superior computational speed compared to, for instance, DeePMD~\cite{MTP_DeePMD_comparison} (while maintaining similar accuracy), no prior studies have combined MTP with PIMD.
    
    In the present work, we combine AL-MTP and PIMD and test the efficiency of this approach. To that end, an interface between i-PI and MLIP-2 was developed. It was used for the calculation of the thermal properties of lithium hydride (LiH) and silicon (Si) systems, in which accounting for NQEs leads to a significant difference from classical molecular dynamics. Specifically, the temperature dependencies of the lattice parameters and thermal expansion (LTE) coefficients were analyzed up to 550 K and 450 K, respectively. The results obtained were compared with quasi-harmonic approximation (QHA) calculations implemented in the phonopy software package \cite{phonopy-phono3py-JPCM, phonopy-phono3py-JPSJ}, experimental data, and PIMD calculations using the universal machine learning force field MatterSim (MS) \cite{yang2024mattersim}. Additionally, partial radial distribution functions for LiH at 50 K were examined. Furthermore, the convergence of results towards the quantum limit as a function of the number of system replicas was investigated for LiH. The developed approach enables a highly accurate description of material properties, taking into account NQEs via the PIMD method, which allowed us to catch negative LTE in Si system.
    
\section{Methods and models}
\label{MAM}
    
\subsection{Path-integral molecular dynamics}

PIMD is a method that takes NQEs into account in calculations of material properties \cite{markland2018_NQE}. PIMD provides an approximate evaluation of the canonical partition function of a system of $N$ quantum particles by representing each particle as a ring polymer of $P$ beads (replicas) connected by harmonic springs. The Hamiltonian of the resulting system is given by:

    \begin{align} \label{Hamiltonian}
        H_P &= H^0_P + V_P,\quad\text{where}
    \\ \label{Kinetic}
        H^0_P &= \sum_{i=1}^{N} \sum_{j = 1}^{P} \Bigg({m_i \omega_P^2 (r_j^{(i)} - r_{j+1}^{(i)})^2 \over 2} + {(p_j^{(i)})^2 \over 2 m_i} \Bigg),
        \quad\text{and}
        \\ \label{Potential}
        V_P &= \sum_{j = 1}^{P} V (r_j^{(1)}, r_j^{(2)}, \dots, r_j^{(N)}).
    \end{align}

    Based on formulas \eqref{Hamiltonian}--\eqref{Potential}, each particle $i$ is represented as a collection of $P$ beads indexed by $j$. For each particle $i$, the periodicity condition $r_1^{(i)} = r_{P +1}^{(i)}$ is valid, and the term responsible for the interaction in \eqref{Kinetic} is a ring of replicas connected by springs with the frequency of $\omega_P = k_B T P / \hbar$. Therefore, for convergence to the quantum limit, the bead number $P$ should be approximately equal to \cite{markland2018_NQE}:
    
    \begin{equation} \label{Replicas_number}
        P \sim \hbar \omega_{\rm max} / k_B T,
    \end{equation}
    where \eqref{Replicas_number} is the ratio of the maximum phonon energy to thermal energy, which quantitatively determines the ``quantumness'' of the system. Each of the $P$ beads of the $N$ particles is described by a potential function $V$ of the system, as can be seen from \eqref{Potential}.
    
    \subsection{Moment tensor potentials}
    
    Additional increase in the required resources due to the presence of replicas of the system, make it inefficient to use quantum-mechanical calculations and, in particular, DFT to describe interactions between atoms. Therefore, to speed up calculations in this work Moment Tensor Potentials (MTPs) implemented in the MLIP-2 software package \cite{novikov2020mlip} were used. MTP is a local potential, i.e. the energy of an atomic configuration is a sum of the energies of local atomic environments of the individual atoms:
    \begin{equation}
        E^{\text{MTP}} = \sum_{i=1}^{N} V({\mathfrak{\boldsymbol n}}_{i}),
    \end{equation}
    where the index $i$ labels $N$ atoms of the system, ${\mathfrak{\boldsymbol n}}_{i}$ denotes the local atomic neighborhood around atom \textit{i} within a certain cut-off radius $R_\text{cut}$, and the function $V$ is the energy of atomic neighborhood: 
    \begin{equation}
        V({\mathfrak{\boldsymbol n}}_{i}) = \sum_{\alpha} \xi_{\alpha} B_{\alpha}({\mathfrak{\boldsymbol n}}_{i}).
    \end{equation}
    where $\xi_\alpha$ are the linear parameters to be fitted and $B_\alpha$ are the basis functions that will be defined below. As fundamental symmetry requirements, all descriptors for atomic environment have to be invariant to translation, rotation, and permutation with respect to the atomic indexing. These requirements are valid for moment tensors descriptors $M_{\mu, \nu}$ and they are used as representations of atomic environments: 
    \begin{equation}
        M_{\mu, \nu}\left({\mathfrak{\boldsymbol n}}_{i}\right)=\sum_j f_\mu\left(\left|r_{i j}\right|, z_i, z_j\right) \underbrace{{\boldsymbol r}_{i j} \otimes \ldots \otimes \boldsymbol{r}_{i j}}_{\nu \text { times }},
    \end{equation}
    where the index $j$ goes through all the neighbors of atom $i$. The symbol ``$\otimes$'' indicates the outer product of vectors, thus ${\boldsymbol r}_{i j} \otimes \cdots \otimes {\boldsymbol r}_{i j}$ is the tensor of rank $\nu$ encoding the angular part. 
    The function $f_\mu$ represents the radial component of moment tensor descriptor:
    \begin{equation}
        f_\mu\left(\left|{\boldsymbol r}_{i j}\right|, z_i, z_j\right)=\sum_{\beta} c_{\mu, z_i, z_j}^{(\beta)} Q^{(\beta)}(|{\boldsymbol r}_{i j}|),
    \end{equation}
    where $z_i$ and $z_j$ denote the atomic species of atoms $i$ and $j$, respectively, ${\boldsymbol r}_{ij}$ describes the positioning of atom $j$ relative to atom $i$, $c_{\mu, z_i, z_j}^{(\beta)}$ are the radial parameters to be fitted and 
    \begin{equation}
        Q^{(\beta)}(|{\boldsymbol r}_{i j}|)=T^{(\beta)}(|{\boldsymbol r}_{i j}|)\left(R_{\text {cut }}-|{\boldsymbol r}_{i j}|\right)^2
    \end{equation}
    are the radial functions consisting of the Chebyshev polynomials $T^{(\beta)}(\left|{\boldsymbol r}_{i j}\right|)$ on the interval $(R_\text{min},  R_\text{cut})$. The number of the radial parameters scales quadratically with the number of atomic species in structures. The descriptors $M_{\mu, \nu}$ with $\nu$ equal to $0, 1, 2, \ldots$ are tensors of different ranks that enable one to define basis functions as all possible contractions of these tensors to a scalar. It is possible to construct an infinite number of $B_{\alpha}$ and in order to restrict this number in the MTP functional form, the level of moment tensor descriptors ${\rm lev}M_{\mu, \nu}$ = 2 + 4$\mu$ + $\nu$ was introduced. If $B_{\alpha}$ is obtained from $M_{\mu_1, \nu_1}$, $M_{\mu_2, \nu_2}$ ,
    $\dots$, then ${\rm lev}B_{\alpha}$ = $(2 + 4\mu_1 + \nu_1)$ + $(2 + 4\mu_2 + \nu_2)$ + $\dots$. Therefore, MTP of level $d$ can be obtained by including a finite number of basis functions with ${\rm lev}B_{\alpha} \leq d$. The total set of parameters to be found is denoted by ${\boldsymbol \theta} = (\{ \xi_{\alpha} \}, \{ c^{(\beta)}_{\mu, z_i, z_j} \})$ and the MTP energy of a structure is denoted by $E^{\text{MTP}} = E({\boldsymbol \theta})$.
    
    \subsection{Extrapolation grade of structures and active learning}\label{extrapolation_grade}
    
    A feature of the MLIP-2 software package is the possibility to both perform training of potentials on independently selected configurations and perform active learning with automatic pre-selection of new configurations. In the second case, in order to construct a training set for MTP fitting, it is necessary to calculate the so-called extrapolation grade of structures. For estimating this grade, the following steps should be performed. Assume that the initial training set contains $K$ structures with energies $E^{\rm DFT}_k$, forces ${\boldsymbol f}^{\rm DFT}_{i,k}$, and stresses $\sigma^{\rm DFT}_{i,k}$, $k = 1, \ldots, K$ calculated with DFT. First, the initial MTP is trained, i.e. by solving the following minimization problem, optimal parameters ${\boldsymbol{\bar{\theta}}}$ are found:
    \begin{equation}
        \begin{array}{c}
            \displaystyle
            \sum \limits_{k=1}^K \Bigl[ w_{\rm e} \left(E^{\rm DFT}_k - E_k({\boldsymbol {\theta}}) \right)^2 + w_{\rm f} \sum_{i=1}^N \left| {\boldsymbol f}^{\rm DFT}_{i,k} - {\boldsymbol f}_{i,k}({\boldsymbol {\theta}}) \right|^2 
            \\
            \displaystyle
            + w_{\rm s} \sum_{i=1}^6 \left| \sigma^{\rm DFT}_{i,k} - \sigma_{i,k}({\boldsymbol {\theta}}) \right|^2 \Bigr] \to \operatorname{min},
        \end{array}
    \end{equation}
    where $w_{\rm e}$, $w_{\rm f}$, and $w_{\rm s}$ are non-negative weights expressing the importance of energies, forces, and stresses in the minimization problem. 
    
    After finding the optimal parameters of ${\boldsymbol{\bar{\theta}}}$, it is necessary to create the following matrix:
    \begin{equation}
        \mathsf{B}=\left(\begin{matrix}
            \frac{\partial E_1\left( {\boldsymbol{\bar{\theta}}} \right)}{\partial \theta_1} & \ldots & \frac{\partial E_1\left( {\boldsymbol{\bar{\theta}}} \right)}{\partial \theta_m} \\
            \vdots & \ddots & \vdots \\
            \frac{\partial E_K\left( {\boldsymbol{\bar{\theta}}} \right)}{\partial \theta_1} & \ldots & \frac{\partial E_K\left( {\boldsymbol{\bar{\theta}}} \right)}{\partial \theta_m} \\
        \end{matrix}\right),
    \end{equation}
    where each row corresponds to a particular structure. The matrix $\mathsf{B}$ is used to construct a subset of structures yielding the most linearly independent rows (physically it means geometrically different structures), which is equivalent to finding a square $m \times m$ submatrix $\mathsf{A}$ of the matrix $\mathsf{B}$ of maximum volume (maximal value of $|{\rm det(\mathsf{A})}|$). To achieve this, the so-called maxvol algorithm is applied \cite{goreinov2010_maxvol}. To determine whether a given structure $\boldsymbol x^*$ obtained during an atomistic simulation is representative, the extrapolation grade $\gamma(\boldsymbol x^*)$ should be calculated, which is defined as
    \begin{equation} \label{Grade}
        \begin{array}{c}
            \displaystyle
            \gamma(\boldsymbol x^*) = \max_{1 \leq j \leq m} (|c_j|), ~\rm{where}
            \\
            \displaystyle
            {\boldsymbol c} = \left( \dfrac{\partial E}{\partial \theta_1} (\boldsymbol{\bar{\theta}}, \boldsymbol x^*) \ldots \dfrac{\partial E}{\partial \theta_m} (\boldsymbol{\bar{\theta}}, \boldsymbol x^*) \right) \mathsf{A}^{-1}.
        \end{array}
    \end{equation}
    This grade specifies the maximal factor by which the determinant $|{\rm det(\mathsf{A})}|$ can increase if ${\boldsymbol x^*}$ is added to the training set. Thus, if the structure $\boldsymbol x^*$ is a candidate for adding to the training set then $\gamma( x^*) \geq \gamma_{\rm th}$, where $\gamma_{\rm th} \geq 1$ is an adjustable threshold parameter which controls the value of permissible extrapolation. Otherwise, the structure does not add new information to the training set.
    
    \subsection{MLIP-2 --- i-PI interface}

    To conduct PIMD simulations we use the i-PI software package \cite{ipi3.0}. This package is based on the principle of server-client interaction via sockets. The server is responsible for the evolution of the nuclear coordinates and starts using i-PI, whereas the calculation of the potential energy, forces and stresses is delegated to one or more instances of an external code, acting as clients. This approach enables the use of different software packages as clients. In addition, when calculations with the system's replicas are performed, the server can distribute information about these replicas among several clients at once using multithreading, which significantly reduces computational time. At the same time, due to the built-in parallelization tools, an developed MLIP-2--i-PI interface \cite{tutorial} can be used to simultaneously run a large number of clients on different cores, which makes PIMD calculations computationally efficient. 
    
\begin{figure}[t]
        \centering
        \includegraphics[width=1\linewidth]{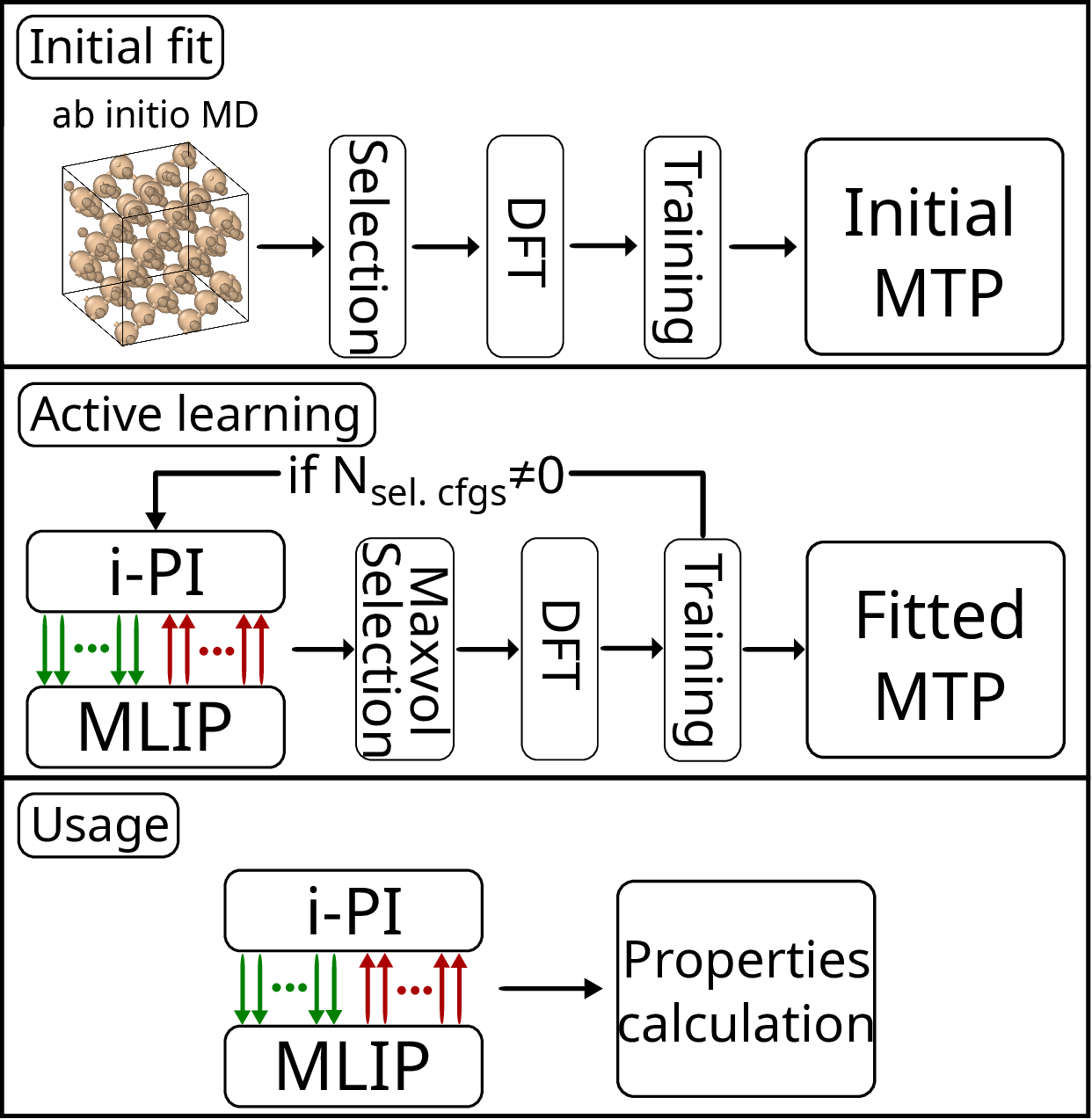}
        \caption{The flowchart describes the general process of potential fitting and its subsequent use in calculating thermal properties using the PIMD method. The upper part of the figure illustrates the fitting of the initial potential. The middle part demonstrates the active learning procedure using the interface between i-PI and MLIP-2. The lower part shows the usage of the fitted potential: the MTP-PIMD calculation with the fitted potential is performed, and the obtained trajectory is used for subsequent processing.}\label{General_scheme}
    \end{figure}

    To actively train potentials in PIMD calculations using the developed interface, the algorithm illustrated in Fig. \ref{General_scheme} was developed. For configuration preselection, two thresholds were introduced: the lower bound $\gamma_{\rm th}$ and the upper bound $\gamma_{\rm cr}$ for the permissible extrapolation. During the PIMD calculation, clients compute the values of $\gamma(x^*)$ for each configuration $x^*$. If $\gamma(x^*) < \gamma_{\rm th}$, the configuration is considered as non-representative and, therefore, is not be added to the training set. When $\gamma_{\rm th} < \gamma(x^*) < \gamma_{\rm cr}$, the configuration is added to a separate set of the preselected configurations for each client, and the PIMD calculation continues. If the value of $\gamma(x^*)$ exceeds $\gamma_{\rm cr}$, the clients stop working, the configuration is added to the preselected set, and, due to the built-in i-PI tools, the server can also be stopped. Afterwards, all the preselected sets are combined, and the matrix $A$ is updated with configurations using the maxvol algorithm. Subsequently, the DFT potential energies, forces, and stresses are calculated, added to the training set, the potential is refitted, and the PIMD calculation is performed again. The active learning procedure is repeated until the number of selected configurations is not equal to zero.
    
    \subsection{Simulation details}
    
    In this work, the developed interface between MLIP-2 and i-PI is used to study the properties of LiH and Si. For each of the considered materials, ab initio molecular dynamics (AIMD) was first performed for 2 ps with a timestep of 1 fs, resulting in a trajectory consisting of 2000 structures. From this trajectory, 40 uncorrelated configurations were selected for more accurate DFT single-point calculations of energies, forces, and stresses in order to form a starting training set and obtain an interpolated potential. Furthermore, some configurations were selected with the maxvol algorithm applied to configurations from the AIMD trajectory. To select configurations during active learning in the PIMD calculations, the lower and upper bounds of the permissible extrapolation were chosen based on the most suitable values for our active learning scheme from the original publication \cite{podryabinkin2017-AL}, with $\gamma_{\rm th} = 2.1$ and $\gamma_{\rm cr} = 10$. It is also worth noting that for training potentials, default values $w_e = 1.0$, $w_f = 0.01$, and $w_s = 0.001$ were chosen as the parameters of the weights of energies, forces, and stresses.

    AIMD, as well as single-point DFT calculations, were performed using Vienna Ab initio Simulation Package (VASP) with the projector-augmented wave method \cite{PAW_Blochl, PAW_VASP}. The Perdew-Burke-Ernzerhof functional \cite{PBE} was used as the exchange-correlation functional. 
    
    To compare the results obtained using the MLIP-2--i-PI interface, we employ MS, a deep learning foundation model for atomic systems across the periodic table, trained over a broad range of temperatures and pressures \cite{yang2024mattersim}. When performing the calculations, the MatterSim-v1.0.0-1M model is used, which is a mini version of the model that allows one to perform the fastest calculations using MS. The Atomic Simulation Environment (ASE) \cite{ase-paper} interface for the i-PI was used to run the MS with the i-PI. 
    
    For calculating the properties of each material, a timestep of 0.5 fs was chosen, and the isotropic barostat with a target pressure value of 0 GPa was employed. The duration of each trajectory was 30 ps at each temperature, with the last 25 ps used to calculate the system's properties. A more detailed description of the computational details for each material is provided below.
    
    \subsubsection{LiH}
    
    For LiH, a $3\times 3\times 3$ cubic supercell consisting of 216 atoms was used. In AIMD, in order to speed up calculations, the cutoff energy $E_{\rm cut}$ was selected to be 300 eV with the $\Gamma$-point only. For single-point DFT calculations, the cutoff energy $E_{\rm cut} = 650$ eV and a k-point mesh of $2 \times 2 \times 2$ were chosen. An MTP of 16-th level with 222 parameters and $R_{\rm min} = 1.42 ~\r{A}$, $R_{\rm cut} = 5 ~\r{A}$ was used. After obtaining the interpolated potential and additional configuration selections from the AIMD trajectory using the maxvol algorithm, the training set consisted of 203 structures. Active learning was carried out in several stages. First, classical MD modeling (with beads number $P=1$) was performed in an NVT ensemble at a temperature of 500 K and a timestep of 0.5 fs. As a result, the training set was increased by 50 configurations. Next, the maximum phonon energy ($\hbar \omega_{\rm max} \approx 0.142 \ \rm eV$) was determined, after which, using formula \eqref{Replicas_number}, the number of replicas $P$ of the system required for performing PIMD calculations at different temperatures was found.  For the better convergence of the results, the number of replicas of the system obtained was multiplied by a factor of 1.7 in all the cases considered. The potential was further trained in an NVT ensemble at a temperature of 100 K with the number of beads $P=30$. Thus, 60 more configurations were selected. The resulting training set included 313 configurations and the potential trained on this training set was used to calculate the dependencies of lattice parameters and thermal expansion coefficients on temperature, as well as radial distribution functions for LiH. 
    
    \subsubsection{Si}

    For Si, a $2 \times 2 \times 2$ cubic supercell consisting of 64 atoms was used. The AIMD calculations employed parameters similar to those for LiH. The values of $E_{\rm cut}$ and the k-point mesh for single-point DFT calculations were 500 eV and $4 \times 4 \times 4$, respectively. In order to obtain most accurate results an MTP of 24-th level with 913 parameters and $R_{\rm min} = 2 ~\r{A}$, $R_{\rm cut} = 5 ~\r{A}$ was used. After obtaining the interpolated potential and additional configuration selections from the AIMD trajectory using the maxvol algorithm, the training set consisted of 115 structures. Active learning was performed according to the same scheme as for LiH. Classical MD modeling (with $P=1$) was conducted in an NVT ensemble at a temperature of 500 K and a timestep of 0.5 fs. As a result, the training set was increased by 235 configurations. Subsequently, the maximum phonon energy ($\hbar \omega_{max} \approx 0.063 \ \rm eV$) was found, $P$ was defined, and the potential was trained in the NVT ensemble at a temperature of 100 K, using a number of beads $P=16$. It is worth noting that the number of configurations selected in this manner is significantly fewer and amounts to only 13 configurations. Thus, the resulting training set included 363 configurations.

    \subsubsection{MTPs validation}

\begin{figure*}[t]
        \centering
        \includegraphics[width=1\linewidth]{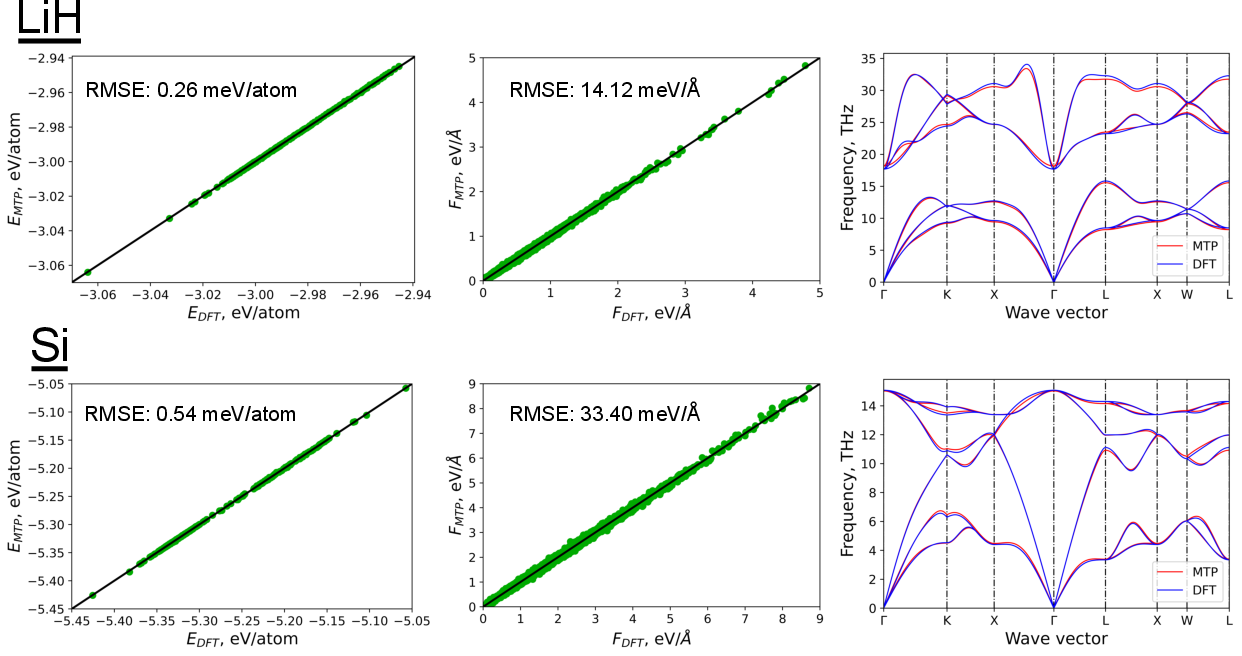}
        \caption{Comparison of DFT-calculated and MTP-calculated energies (on the left) and forces (in the middle) of structures from the training set, as well as phonon band structures (on the right) for the LiH and Si systems.}\label{MTP_validation}
    \end{figure*}

    To validate the potentials obtained during the training procedure, a comparison of the DFT-calculated and MTP-calculated energies of structures from the training set and forces, as well as phonon band structures, was performed. The resulting dependencies are presented in Fig. \ref{MTP_validation}. Potentials errors are presented in Table \ref{RMSE_1}.

    \begin{table}[t]
    \centering
    \begin{tabular}{l c c c}
    \hline
      {} & Train RMSE & Train RMSE & Train RMSE \\
      {} & on energy & on forces & on stresses  \\
      {} & (meV/atom) & (meV/\r{A}) & (MPa) \\
    \hline
      LiH & 0.26 & 14.12 & 61.75 \\
    \hline
      Si & 0.54 & 33.4 & 71.61 \\
    \hline
    \end{tabular}
    \caption{The root mean square errors of the potentials for energies, forces, and stresses}\label{RMSE_1}
    \end{table}

    \section{Results and discussion}
    
    \subsection{Quantum limit convergence}
    
    \begin{figure}[t]
        \centering
        \includegraphics[width=1\linewidth]{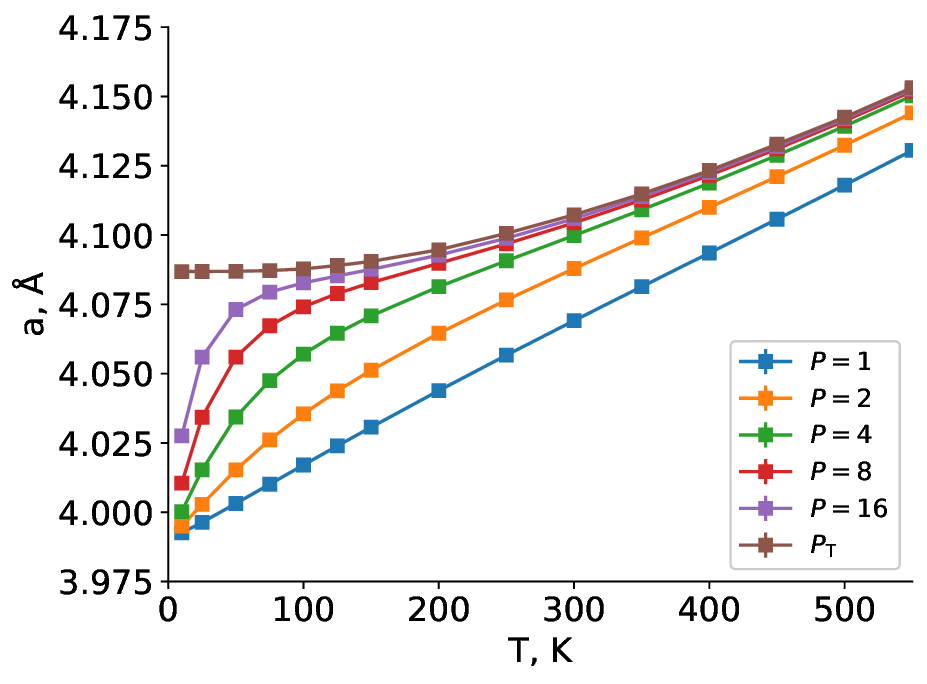}
        \caption{The dependence of the lattice parameter $a$ on temperature for different values of the parameter $P$ (number of copies of the system) for the LiH system. The $P_{\rm T}$ curve corresponds to the value of the number of replicas of the system $P = 1.7 \cdot \hbar \omega_{\rm max} /k_B T$.}\label{fig2}
    \end{figure}

    The study of the effect of NQEs on the thermal properties of materials began with the search for the optimal value of the number of system replicas $P$ using the example of the LiH system. For this purpose, a study of the temperature dependence of the lattice parameter $a$ was first performed. Calculations were completed for $P = 1, 2, 4, 8, 16$, and also for the value of $P$ obtained by the formula \eqref{Replicas_number} and multiplied by a factor of 1.7. The resulting dependence is illustrated in Fig. \ref{fig2}. It can be seen that with an increase in the parameter $P$, under the influence of zero-point oscillations, the approximately linear dependence obtained in the purely classical case ($P=1$) shifts to the region of higher values of $a$, i.e., the renormalization of the lattice parameters occurs.

    The greatest difference in the lattice parameters is observed at a temperature of 0 K: at $P=1$ the lattice parameter is $3.990 \r{A}$, and at $P = P_0$ it is equal to $4.084 \r{A}$. Therefore, the change in the lattice parameter is $0.094 \r{A}$, which is about 2\% of $a(0)$. With the increase in temperature, the influence of NQEs decreases. This is manifested in a decrease in the difference in the values of $a$, as well as the fact that fewer replicas of the system are required to obtain a ``quantum'' system. Thus, starting from 100 K, 16 beads are sufficient for accuracy calculation of $a(T)$, 8 beads are sufficient at 200 K, and 4 beads are sufficient at 400 K. This matches well with the values of $P$ obtained using the formula \eqref{Replicas_number} and equal to 16.6, 8.3, and 4.1, respectively. Therefore, to obtain accurate results with minimal computational resources, it is sufficient to use $P$ replicas of the system defined by formula \eqref{Replicas_number}; however, we recommend using a slightly higher value of $P$ to obtain the most accurate results.

    This conclusion is also confirmed by the behavior of radial distribution functions (RDFs), which also show the influence of NQEs on the materials properties. The formula for calculating RDFs can be written as follows \cite{levine2011fast}:

    \begin{equation}
        \label{RDF}
        g(r) = \underset{dr \to 0}{\rm{lim}} {p (r) \over 4 \pi (N_{\rm pairs} / V) r^2 dr} \ ,
    \end{equation}
    where $r$ is the distance between a pair of particles, $p(r)$ is the average number of atom pairs found at a distance between $r$ and $r + dr$, $V$ is the total volume of the system, and $N_{\rm pairs}$ is the number of unique pairs of atoms. 

    \begin{figure}[t]
        \centering
        \includegraphics[width=1\linewidth]{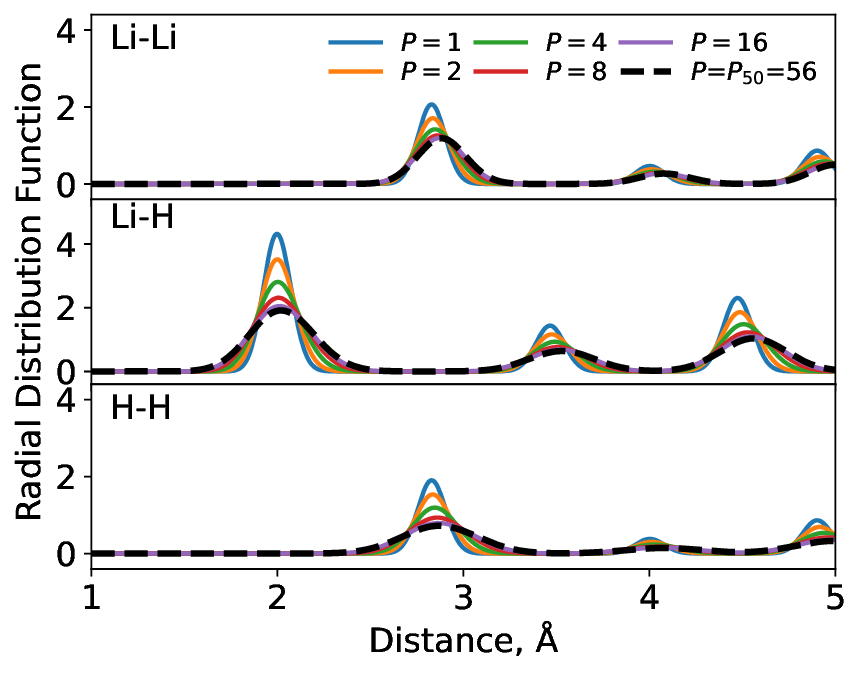}
        \caption{Partial lithium-lithium, lithium-hydrogen, and hydrogen-hydrogen radial distribution functions for the LiH system at 50 K for a different number of replicas of the system. The dashed line corresponds to the optimal beads number $P$ for the chosen temperature.}\label{fig4}
    \end{figure}
    
    According to the \eqref{RDF}, for the trajectories obtained during the calculation of $a(T)$ with different parameters $P$, the partial RDFs for pairs of lithium-lithium, lithium-hydrogen, and hydrogen-hydrogen atoms were calculated. For example, the RDF shown in Fig. \ref{fig4} was obtained at 50 K. It can be seen that the curve for the classical case has narrow and well-defined peaks that characterize the behavior of the crystal structure. As a result of the increase in the number of replicas of the system and the inclusion of NQEs, the peaks are widened and the distribution function becomes blurred. It is worth noting that the RDF converges to the limiting value of the $P_{\rm T}$ curve, which is clearly shown in Fig. \ref{fig4}.

    \subsection{LiH}
    
    \begin{figure*}[t]
        \centering
        \includegraphics[width=0.49\linewidth]{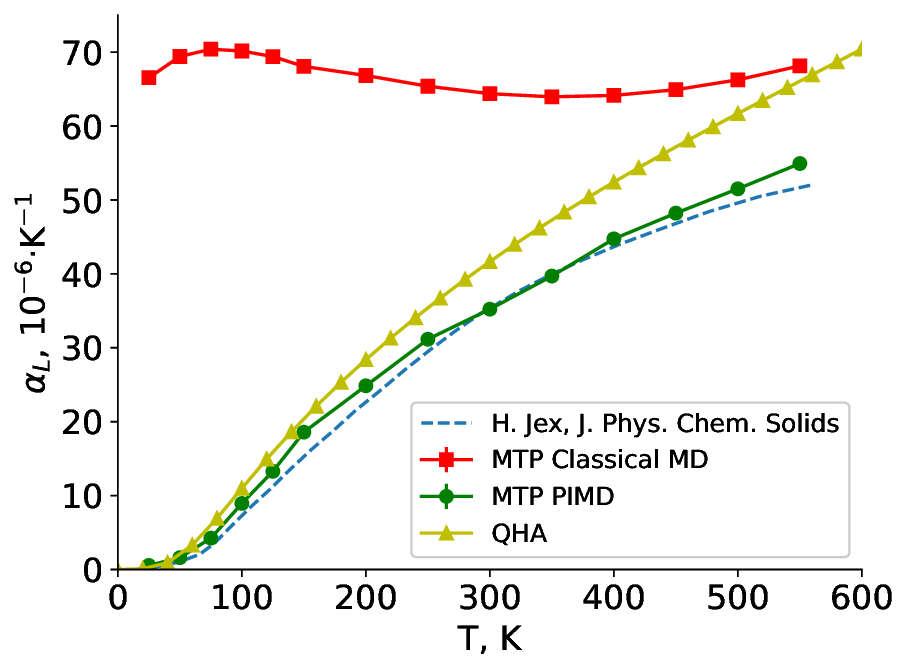}
        \includegraphics[width=0.49\linewidth]{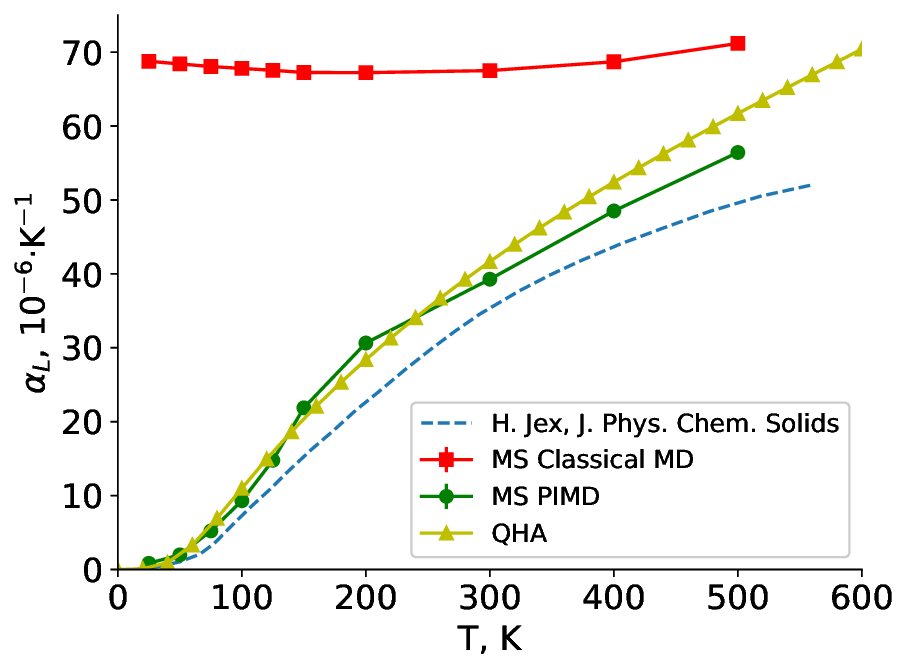}
        \caption{The dependence of the linear LTE coefficient $\alpha_{\rm L}$ for the LiH system on temperature for the cases of classical MD (red squares) and PIMD (green circles) using MTP (on the left) and MS (on the right), QHA (yellow triangles), and the experimental data obtained in the article \cite{LiHLTE} (blue dashed line).}\label{fig3}
    \end{figure*}
    
    Based on the dependence of the lattice parameter $a$ on temperature, another thermal property can be calculated, namely, the coefficient of the linear lattice thermal expansion (LTE):
    \begin{equation}
        \alpha_{\rm L} = {1 \over a_0} {\partial a \over \partial T} \, ,
    \end{equation}
    where $\alpha_{\rm L}$ is the linear LTE coefficient, $a_0$ is the lattice parameter at 0 K. The result of calculating $\alpha_{\rm L}$ for classical MD and PIMD for MTPs and MS applications is shown in Fig. \ref{fig3}. 

    As can be seen from the figures obtained, there is a significant difference between the classical and ``quantum'' cases. Thus, for classical MD, $\alpha_{\rm L}(T)$ values are approximately constant and located significantly higher than for the case of PIMD at temperatures up to 500 K and more. This LTE behavior can primarily be attributed to the nearly linear growth of $a(T)$ at $P = 1$ which is shown in Fig. \ref{fig2}. Thus, classical MD modeling is applicable only for temperatures at which the zero-point energy is comparable and lower than the thermal energy.

    At the same time, the results obtained by the PIMD method demonstrate a high degree of agreement with the experimental data presented in \cite{LiHLTE} and QHA. It is worth noting that both applied potentials qualitatively show the correct behavior of the dependence of $\alpha_{\rm L}$ on temperature. The MTP-PIMD approach allowed us to obtain values that almost coincide with the experimental ones, while the MS-PIMD approach showed slightly overestimated values. The results of QHA are also in good agreement with the results of PIMD modeling, however, already at temperatures of about 150 K, LTE computed in the scope of the QHA overestimates experimental results, which is associated with the increasing role of anharmonicity of the potential energy surface with increasing temperature.
    
    \subsection{Si}

    For the Si system, PIMD simulations showed that, similar to the results obtained for LiH, NQEs depend on temperature and become more pronounced as it decreases. It was found that the difference in the lattice parameters for the classical and ``quantum'' cases is an order of magnitude smaller and equal to 0.009 $\r{A}$, which is about 0.2\% of $a(0)$ and indicates a decrease in the effect of NQEs with increasing atomic mass of the system.
    
    \begin{figure*}[t]
        \centering
        \includegraphics[width=0.49\linewidth]{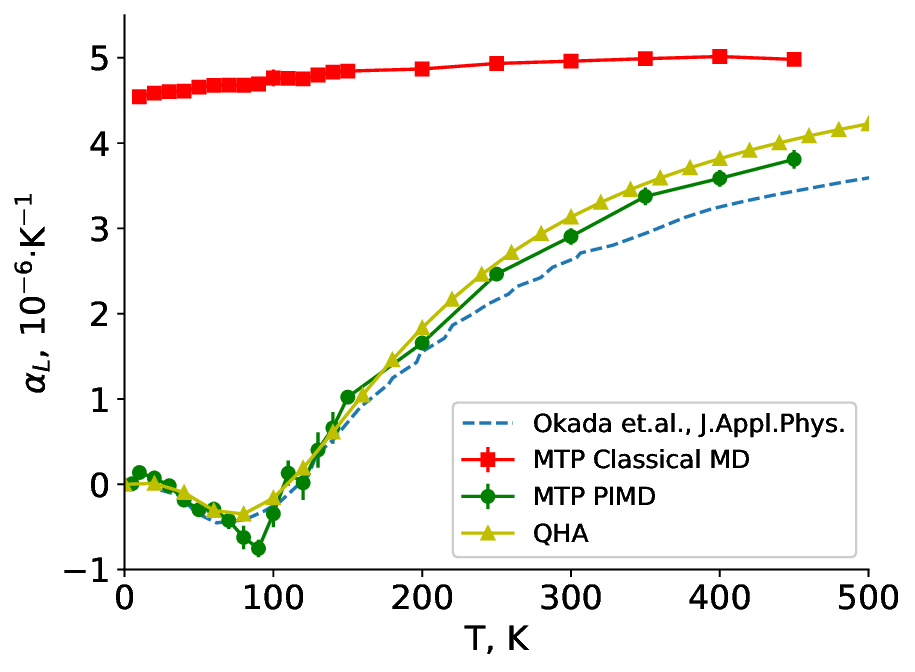}
        \includegraphics[width=0.49\linewidth]{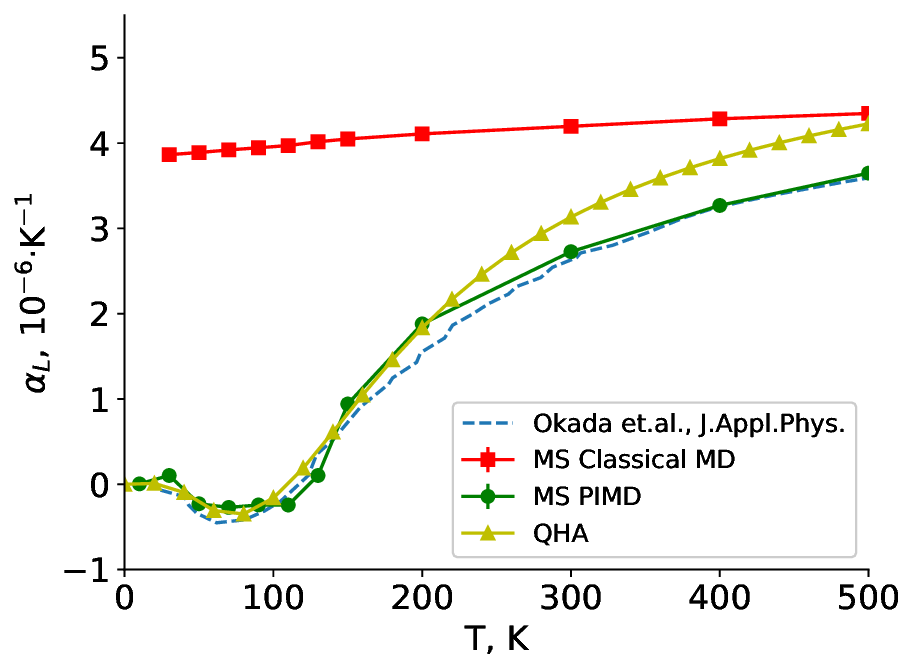}
        \caption{The dependence of the linear LTE coefficient $\alpha_{\rm L}$ for the Si system on temperature for the cases of classical MD (red squares) and PIMD (green circles)  using MTP (on the left) and MS (on the right), QHA (yellow triangles), as well as experimental data obtained in the article \cite{SiLTE} (blue dashed line).}\label{fig7}
    \end{figure*} 

    Despite this, there is a noticeable difference between classical MD and PIMD for linear LTE coefficients, which, as for the LiH system, manifests itself in a higher and approximately constant value of $\alpha_{\rm L}(T)$ for classical MD. As can be seen from the figures obtained, the key feature of the Si system is the presence of a negative LTE coefficient (volume reduction during heating) in the temperature range from about 30 K to 110 K. This anomalous behavior of $\alpha_{\rm L}$ was discovered and studied in the following works \cite{SiLTE, Si_LTE_2, Si_LTE_3}. It is worth noting that both approaches used (MTP-PIMD and MS-PIMD) made it possible to detect this effect and obtain high agreement with the experimental data presented in \cite{SiLTE}. The MTP-PIMD approach, despite the correct description of LTE behavior in the specified temperature range, gives slightly higher values of $\alpha_{\rm L}(T)$ at temperatures above 200 K, while for MS-PIMD the values almost exactly match the experimental data. QHA calculations also allowed us to show the presence of abnormal LTE coefficient behavior and high consistency with PIMD modeling, however, at temperatures above 200 K, due to the increased role of anharmonicity in the system, QHA gives overestimated values of $\alpha_{\rm L}(T)$.
    
    \section{Conclusions}

    In this study, we combine PIMD with MTPs and MS  to efficiently and accurately account for NQEs. This approach was applied to investigate the properties of LiH and Si, as well as to analyze the differences between classical and quantum nuclear dynamics.

    To enable MTP-PIMD calculations, we developed an interface between MLIP-2 and i-PI, allowing MLIP-2 to function as a client for i-PI. The high efficiency of the interface is ensured by the possibility of both active learning of potentials and the parallel creation of several clients on different cores to reduce computing time. In addition, using the MTP-PIMD approach significantly reduces the time spent on calculations compared to performing DFT-PIMD calculations. For example, if the DFT-PIMD approach requires more than $10^8$ CPU-hours for the Si system, then, due to the very high speed of performing calculations with MTP potentials \cite{MLIPs_performance}, the MTP-PIMD approach requires only about $10^4$ CPU-hours (including AIMD and active learning of the potential), which is a feasible task for modern high-performance computing systems.

    The results obtained demonstrate the importance of accounting for NQEs in accurate modeling the thermal expansion of materials and in studying the behavior of their radial distribution functions. For the LiH system we investigated how the number of replicas of the system used in PIMD modeling affects its properties. It was shown that at low temperatures, the values of the lattice parameter increase compared to the classical case, and that during the transfer to the ``quantum'' system, the peak widths of the radial distribution functions increase.

    Additionally, it was found that to simulate the quantum behavior of systems, it is sufficient to use $\mathrm P = P_{\rm T} = \hbar \omega_{\rm max} / k_B T$ replicas of the system in calculations, which in some cases significantly reduces the required computing resources. For Si, it was shown that accounting for NQEs enables us to observe abnormal behavior in the linear thermal expansion (LTE) coefficient as temperature changes.

    The developed methodology and the interface opens up a wide range of possibilities for studying the properties of various materials in which NQEs play an important role. Thus, the developed approach made it possible to obtain a high degree of consistency of the MTP-PIMD results with experimental data, as well as MS-PIMD calculations for the considered systems.

    \section{Acknowledgments}

    This work was supported by the Russian Science Foundation (grant number 23-13-00332).

    \section{Conflict of interest}

    The authors have no conflicts of interest to disclose.

    \section{Data availability}

    We prepared a tutorial available at the link \cite{tutorial}, which describes in detail the operation of the developed interface between MLIP-2 and i-PI. The tutorial provides examples of active learning of MTPs in PIMD calculations and RDF calculations for LiH, as well as calculations of linear LTE coefficients for Si.

    The source code of MLIP-2 can be obtained from \cite{mlip2-source}.
    
    \bibliographystyle{elsarticle-num} 
    \bibliography{MLIP-PIMD}
    
\end{document}